\documentclass[article,preprint,showkeys,secnumarabic,amsfonts,showpacs,amsmath,amssymb]{revtex4}
\usepackage{amsmath}
\usepackage{appendix}
\usepackage[dvips]{color}
\usepackage{epsfig}
\usepackage{graphicx}
\usepackage{amssymb,epsf}
\usepackage{epstopdf}
\usepackage{makeidx}
\usepackage{verbatim}
\usepackage{hyperref}
\usepackage{multirow}
\usepackage{natbib}
\usepackage{amsthm} 
\usepackage{cancel} 
\usepackage{enumitem}
\usepackage{mathrsfs} 
\usepackage[utf8]{inputenc}
\usepackage{amsmath, amssymb, amsthm}
\usepackage{tikz} 
\usetikzlibrary{arrows.meta}

\newcommand{\Sd}{S_{\delta}}
\newcommand{\rh}{r_{h}}

\begin{document}
	
	\title{A smooth BTZ black bounce with an extremal null throat}
	
	\author{Farzad Milani}\email{fmilani@tvu.ac.ir}\affiliation{\small{Department of Basic Sciences, Technical and Vocational University (TVU), Tehran, Iran.}}
	\date{\today}
	
	\begin{abstract}
		We study a static, circularly symmetric deformation of the non-rotating BTZ black hole obtained by inserting a smooth transition function into the \emph{inverse} radial metric component, $g^{rr}=S_\delta(r)F(r)$ with $S_\delta=\tanh[(r-r_h)/\delta]$, leaving $g_{tt}=-F$ untouched. This was motivated by the proposal that such a construction realizes a Lorentzian-to-Riemannian signature change at the horizon; we show that it does not. In coordinates $r-r_h=q^2$ with an advanced time, the metric extends real-analytically across $r=r_h$, and the extension is Lorentzian: $q=0$ is a regular null hypersurface, a degenerate Killing horizon with vanishing surface gravity, beyond which lies a second, isometric copy of the exterior. The areal radius has a minimum there, so the geometry is a black bounce; the would-be Riemannian branch is a separate geometry the Lorentzian sector never reaches. We give the effective source in closed form, an invariant account of the energy conditions, and identify the near-throat geometry as AdS$_2\times S^1$. The scalar effective potential is proven strictly positive for every mode, and the throat circle is a minimal surface whose length gives an entropy $\pi r_h/2G$, reproduced independently by the Wald--Noether charge and by a Cardy estimate from the computed Brown--York mass -- concordant results for which no first law is available since $\kappa=0$. The throat carries an Aretakis-type instability, with a conserved leading transverse derivative and a linearly growing subleading one. We also record a negative result: smoothing $g_{tt}$ instead, as in the Lorentzian-Euclidean Schwarzschild proposal, is singular at the horizon for any finite smoothing width. We state explicitly what the construction does not establish.
	\end{abstract}
	
	
	
	\maketitle
	
	\section{Introduction}\label{sec:intro}
	
	Signature-changing metrics have been studied in quantum cosmology and in classical relativity for several decades~\cite{HartleHawking,GibbonsHartle,HalliwellHartle,EllisEtAl,Ellis92,DrayEtAl}. Capozziello, De Bianchi and Battista~\cite{CDB} brought the idea to black holes, proposing a ``Lorentzian--Euclidean'' Schwarzschild geometry in which the discontinuous sign function $\varepsilon(r)$ multiplies $g_{tt}$, so that the region inside the horizon acquires an ultrahyperbolic signature, and interpreting the resulting kinematics as an ``atemporality'' mechanism preventing access to the singularity. The distributional curvature at the change surface is handled by a Hadamard \emph{partie finie} prescription. Bartolo, Caponio, Germinario and S\'anchez~\cite{Bartolo} subsequently analysed such transitions geometrically, distinguishing degeneracy of the metric $g$ from degeneracy of the dual metric $g^{*}$---the causal cones collapsing to a line in the first case and to a hyperplane in the second---and clarified that the radial free-fall proper time in Ref.~\cite{CDB} is in fact finite.
	
	This paper began as an attempt to realize the same idea smoothly in the BTZ geometry~\cite{BTZ,BHTZ}, by replacing the sign function with a $\tanh$ profile~\cite{Milani}. Two results emerged, and neither is the one we set out to obtain.
	
	First, a genuinely negative result. If the transition function multiplies $g_{tt}$, i.e.\ $g_{tt}=-\Sd F$ with $g_{rr}=1/F$---the direct smooth transcription of the ansatz of Ref.~\cite{CDB} to BTZ---then near the horizon $g_{tt}$ has a double zero while $g_{rr}$ has a simple pole, and the Ricci scalar diverges,
	\begin{equation}
		R\;\simeq\;-\frac{F'(\rh)}{x},\qquad x=r-\rh\to0,
		\label{eq:negative}
	\end{equation}
	for \emph{every} finite smoothing width $\delta$. (Numerically, at $M=\ell=\rh=1$ and $\delta=0.1$: $R=1989$ at $x=10^{-3}$ against the asymptote $2000$.) Smoothing does not regularize this class of ansatz; the distributional treatment of Ref.~\cite{CDB} is a necessity of the construction rather than a technical convenience.
	
	Second, the alternative of placing the transition function in the inverse radial component,
	\begin{equation}
		ds^{2}=-F(r)\,dt^{2}+\frac{dr^{2}}{\Sd(r)F(r)}+r^{2}d\phi^{2},
		\label{eq:metric}
	\end{equation}
	with
	\begin{equation}
		F=-M+\frac{r^{2}}{\ell^{2}},\qquad \Sd=\tanh\!\Big(\frac{r-\rh}{\delta}\Big),
	\end{equation}
	does produce finite curvature invariants at $r=\rh$. Formally, for $r<\rh$ both $\Sd$ and $F$ are negative, so $g_{rr}>0$ and $g_{tt}=-F>0$, and one is tempted to read the interior as Riemannian and $r=\rh$ as a signature-change surface. That reading is incorrect, and establishing why is the main content of this paper. The chart~\eqref{eq:metric} is singular at $\rh$ ($g_{rr}$ has a double pole, $\det g\to\infty$), and it is the \emph{dual} metric that degenerates there, in the sense of Ref.~\cite{Bartolo}. In coordinates adapted to that degeneracy the geometry is perfectly regular, and its unique analytic extension is Lorentzian: the horizon is a regular null hypersurface with vanishing surface gravity, and beyond it lies a second copy of the exterior region. The areal radius bounces. The geometry belongs to the black-bounce family of Simpson and Visser~\cite{SimpsonVisser,LoboEtAl}, of which a BTZ member is already known~\cite{FurtadoAlencar}; what is specific here is that the bounce occurs exactly at the horizon, which is consequently degenerate.
	
	Section~\ref{sec:extension} constructs the extension and establishes its uniqueness, the null character of the throat and the second sheet. Section~\ref{sec:curvature} gives the curvature and the effective source in closed form; Sec.~\ref{sec:ec} the energy conditions. Section~\ref{sec:geo} treats geodesics in the extension and establishes analytic regularity of the crossing for the whole family of conserved charges. Section~\ref{sec:throat} identifies the near-throat AdS$_{2}\times S^{1}$ geometry, computes the entropy by three independent routes and the quasilocal mass, and states what can and cannot be said thermodynamically. Section~\ref{sec:scalar} proves positivity of the scalar effective potential for all angular modes and derives the Aretakis-type behaviour at the throat. Section~\ref{sec:scope} states the scope and limitations explicitly, and Sec.~\ref{sec:concl} concludes. Detailed derivations of all boxed results are collected in Appendices~\ref{app:A}--\ref{app:E}. We use $G=c=1$, take $\phi\in[0,2\pi)$, and note that $M$ is dimensionless in these conventions, with $\rh=\sqrt{M}\,\ell$.
	
	\section{The extension, the throat, and the second sheet}\label{sec:extension}
	
	\subsection{The chart \eqref{eq:metric} is singular at the horizon}
	
	Near $\rh$, $F\simeq F'(\rh)x$ and $\Sd\simeq x/\delta$, so $g^{rr}=\Sd F$ has a \emph{double} zero and
	\begin{equation}
		g_{rr}\simeq\frac{\delta}{F'(\rh)\,x^{2}}\to\infty,\qquad \det g=-\frac{r^{2}}{\Sd}\to\infty .
	\end{equation}
	Consequently the proper radial distance to the horizon diverges logarithmically, $\int dr/\sqrt{\Sd F}\simeq\sqrt{\delta/F'(\rh)}\,|\ln x|$, from both sides, and the Killing time along an infalling geodesic diverges as $x^{-1/2}$ (numerically $t=10.3,\,100.3,\,1000.3$ at $x=10^{-3},10^{-5},10^{-7}$), while the proper time converges. None of this by itself indicates a singularity: exactly the same features occur at an extremal Reissner--Nordstr\"om horizon, which is a perfectly regular null hypersurface. What it does indicate is that~\eqref{eq:metric} is a bad chart there, and that the question of what lies at $r=\rh$ must be settled in a chart adapted to the degeneracy.
	
	\subsection{A regular chart and the analytic extension}
	
	Set
	\begin{equation}
		r=\rh+q^{2},\qquad dv=dt+\frac{dr}{F\,\sigma},\qquad \sigma^{2}=\Sd,
		\label{eq:coordchange}
	\end{equation}
	where $\sigma$ is the \emph{signed} root. Writing $\tanh y=y\,g(y)$ with $g$ analytic and $g(0)=1$, one has $\sigma=q\sqrt{g(q^{2}/\delta)/\delta}$, which is analytic and odd in $q$ (Appendix~\ref{app:A}); at $M=\ell=\rh=1$, $\delta=0.1$ it reads $\sigma=\sqrt{10}\,q-\tfrac{50\sqrt{10}}{3}q^{5}+O(q^{7})$. The metric becomes
	\begin{equation}
		ds^{2}=-F(\rh+q^{2})\,dv^{2}+\beta(q^{2})\,dv\,dq+(\rh+q^{2})^{2}d\phi^{2},
		\label{eq:extension}
	\end{equation}
	with
	\begin{equation}
		\beta=\frac{2\,dr/dq}{\sigma}=4\sqrt{\frac{\delta}{g(q^{2}/\delta)}},\qquad \beta(0)=4\sqrt{\delta}\neq0 .
	\end{equation}
	Every coefficient is analytic in $q$, and the determinant of the $(v,q)$ block is $-\beta^{2}/4\to-4\delta<0$: the metric is finite, Lorentzian and non-degenerate at $q=0$. The extension to $q<0$ is therefore available, and since the metric is real-analytic it is the \emph{unique} analytic extension.
	
	Two structural facts follow immediately.
	
	\emph{(i) The throat is null.} From~\eqref{eq:extension}, $g^{qq}=4F/\beta^{2}$, which vanishes at $q=0$ because $F(\rh)=0$. Hence $\Sigma:q=0$ is a null hypersurface. It is a Killing horizon of $\partial_{v}$, and since $g_{vv}=-F$ vanishes quadratically in $q$, it is \emph{degenerate}: the surface gravity is
	\begin{equation}
		\kappa=\tfrac12 F'(\rh)\sqrt{\Sd(\rh)}=0 \qquad\text{for every }\delta>0 .
		\label{eq:kappa}
	\end{equation}
	
	\emph{(ii) Beyond the throat lies a second exterior.} For $q<0$ one has $r=\rh+q^{2}>\rh$ again. The coefficients in~\eqref{eq:extension} are even functions of $q$, so the line element is invariant under $(v,q)\to(-v,-q)$ and the two sheets are isometric. On both sheets $F>0$, so $\partial_{v}$ is timelike and there is no trapped region; the areal radius has a \emph{minimum} at $\Sigma$. The geometry is a black bounce with a degenerate null throat.
	
	The Riemannian branch $r<\rh$ of~\eqref{eq:metric} is thus not reached by the extension of the Lorentzian sector. It is a separate geometry, and we make no claims about it here beyond noting, for completeness, that with the $\tanh$ profile its centre is singular: $\Sd'(0)=\mathrm{sech}^{2}(\rh/\delta)/\delta\neq0$ for every finite $\delta$, and the Ricci scalar of that branch behaves as $M\Sd'(0)/r$ as $r\to0$. A profile with $\Sd'(0)=0$ would be required for a regular centre; the exponentially small coefficient makes this defect easy to miss numerically. Since the Lorentzian spacetime never reaches $r<\rh$, this does not affect anything below.
	
	\begin{figure}[t]
		\includegraphics[width=\columnwidth]{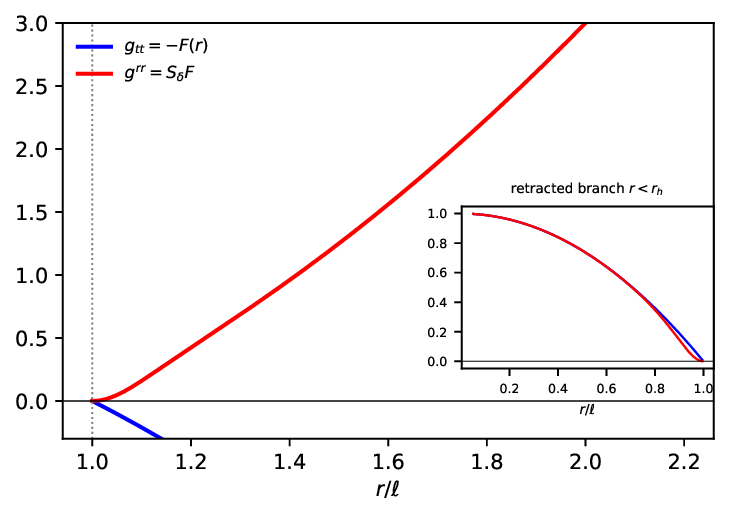}
		\caption{Exterior structure at $M=\ell=1$, $\delta=0.1$. The double zero of $g^{rr}=\Sd F$ at $\rh$ is the double pole of $g_{rr}$; this is a degeneracy of the \emph{dual} metric in the sense of Ref.~\cite{Bartolo}, not of the metric. Inset: the retracted branch $r<\rh$, shown for visual comparison only. That branch is \emph{not} part of the Lorentzian spacetime discussed in this paper (Sec.~\ref{sec:extension}).}
		\label{fig:structure}
	\end{figure}
	
	\begin{figure}[t]
		\includegraphics[width=\columnwidth]{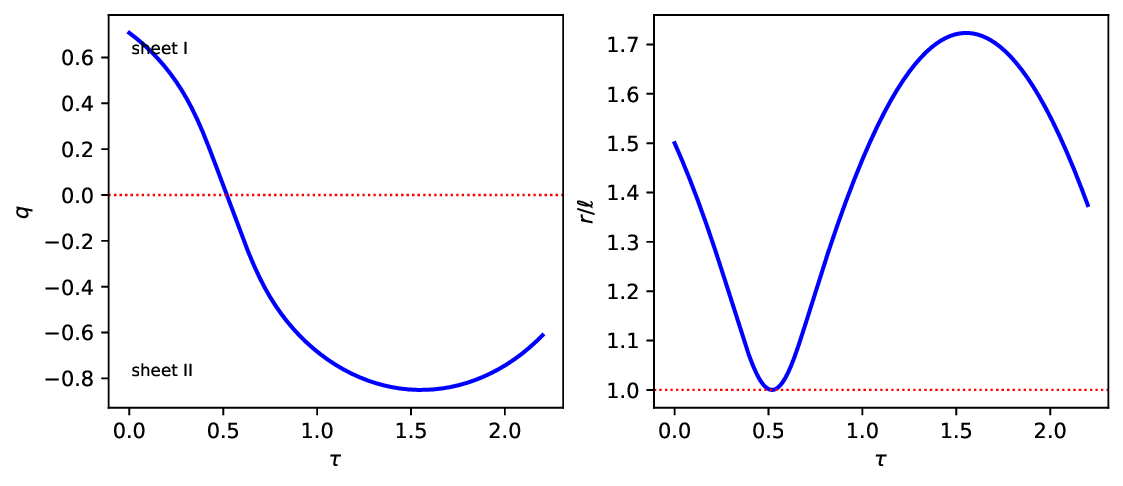}
		\caption{A radial timelike geodesic integrated as a full system in the regular chart~\eqref{eq:extension}. Left: $q$ passes smoothly through $0$, from the first sheet to the second. Right: the areal radius attains its minimum $\rh$ and increases again---the bounce. The norm is conserved to six digits across the throat.}
		\label{fig:extension}
	\end{figure}
	
	\section{Curvature and the effective source}\label{sec:curvature}
	
\begin{figure}[t]
		\includegraphics[width=\columnwidth]{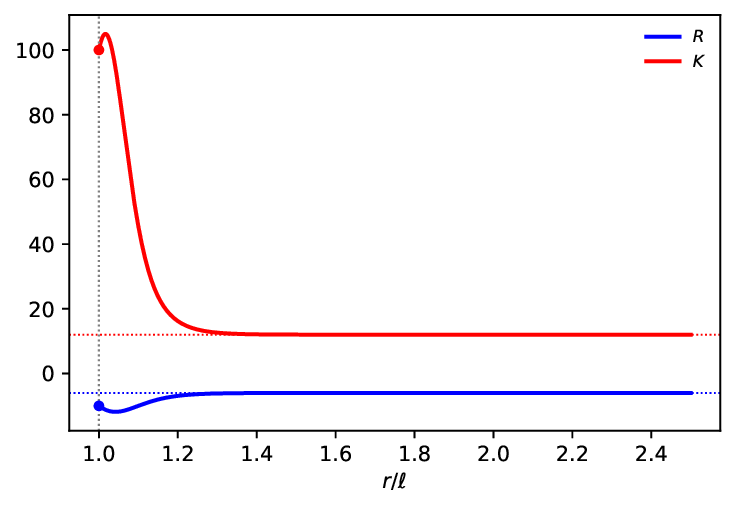}
		\caption{Ricci and Kretschmann scalars on the Lorentzian sector $r\geq\rh$, for $M=\ell=1$, $\delta=0.1$. Dotted lines mark the BTZ asymptotic values $-6/\ell^{2}$ and $12/\ell^{4}$; the markers at $r=\rh$ are the closed-form throat values~\eqref{eq:inv}.}
		\label{fig:invariants}
	\end{figure}

	On the Lorentzian sector the mixed Ricci components have finite limits at $\Sigma$ for every $\delta>0$,
	\begin{equation}
		R^{t}{}_{t}(\Sigma)=R^{r}{}_{r}(\Sigma)=-\frac{F'(\rh)\Sd'(\rh)}{4},\qquad R^{\phi}{}_{\phi}(\Sigma)=0,
	\end{equation}
	so that, using the three-dimensional identity $K=4R_{\mu\nu}R^{\mu\nu}-R^{2}$ (Appendix~\ref{app:B}),
	\begin{equation}
		R(\Sigma)=-\frac{\rh}{\ell^{2}\delta},\qquad K(\Sigma)=\frac{\rh^{2}}{\ell^{4}\delta^{2}} .
		\label{eq:inv}
	\end{equation}
	The same values are obtained in the regular chart~\eqref{eq:extension}, where $R$ is manifestly finite at $q=0$. Asymptotically $R\to-6/\ell^{2}$, the exterior approaching BTZ exponentially fast (Fig.~\ref{fig:invariants}). Note that~\eqref{eq:inv} diverges as $\delta\to0$: the smooth family does not converge to BTZ in curvature, and $\delta$ cannot be removed at the end of the calculation (see Sec.~\ref{sec:scope}).
	
	In three dimensions the Weyl tensor vanishes, so any metric that is not of constant curvature is necessarily sourced. Defining
	\begin{equation}
		8\pi T^{\mu}{}_{\nu}\equiv G^{\mu}{}_{\nu}+\Lambda\delta^{\mu}{}_{\nu},\qquad \Lambda=-\frac{1}{\ell^{2}},
	\end{equation}
	which is a \emph{definition}, not the solution of any field equation, and carries no implication that a matter model producing it exists, one finds for a generic smooth transition function $S(r)$ (Appendix~\ref{app:B})
	\begin{align}
		8\pi T^{t}{}_{t}&=\frac{FS'}{2r}+\frac{S-1}{\ell^{2}},\nonumber\\
		8\pi T^{r}{}_{r}&=\frac{S-1}{\ell^{2}},\label{eq:source}\\
		8\pi T^{\phi}{}_{\phi}&=\frac{rS'}{2\ell^{2}}+\frac{S-1}{\ell^{2}} .\nonumber
	\end{align}
	Conservation $\nabla_{\mu}T^{\mu\nu}=0$ holds automatically by the Bianchi identity and is not an independent check.
	
	\section{Energy conditions}\label{sec:ec}
	
	On the Lorentzian sector, with $\rho=-T^{t}{}_{t}$, $p_{r}=T^{r}{}_{r}$, $p_{\phi}=T^{\phi}{}_{\phi}$,
	\begin{equation}
		8\pi(\rho+p_{r})=-\frac{FS'}{2r},\qquad 8\pi(\rho+p_{\phi})=\frac{MS'}{2r} .
		\label{eq:nec}
	\end{equation}
	The angular null energy condition is satisfied everywhere. The radial one is violated wherever $F>0$ and $S'>0$---that is, throughout the exterior, with magnitude peaking at $\approx0.40$ near $x\approx\delta$ and decaying exponentially. We emphasise that the violation is \emph{not} compactly supported: $S'>0$ at every finite $r$, so the tails never terminate (e.g.\ $8\pi(\rho+p_{r})=-6.2\times10^{-8}$ at $x=\ell$). Integrated against the invariant proper volume $2\pi r\,d\ell$, the total radial NEC violation is $-1.58$ at the reference parameters; the corresponding coordinate-measure integral, $-0.066$, is not a geometric quantity and should not be quoted. The weak energy condition is violated over part of the shell ($8\pi\rho$ dips to $-0.16$); the dominant energy condition fails wherever the radial NEC does, since $p_{r}<0$ there; the strong energy condition is satisfied throughout the shell. Localized energy-condition violation of this kind is characteristic of bouncing and regular-interior constructions~\cite{SimpsonVisser,LoboEtAl,Dymnikova,HaywardRBH}.
	
	\begin{figure*}[t]
		\includegraphics[width=\textwidth]{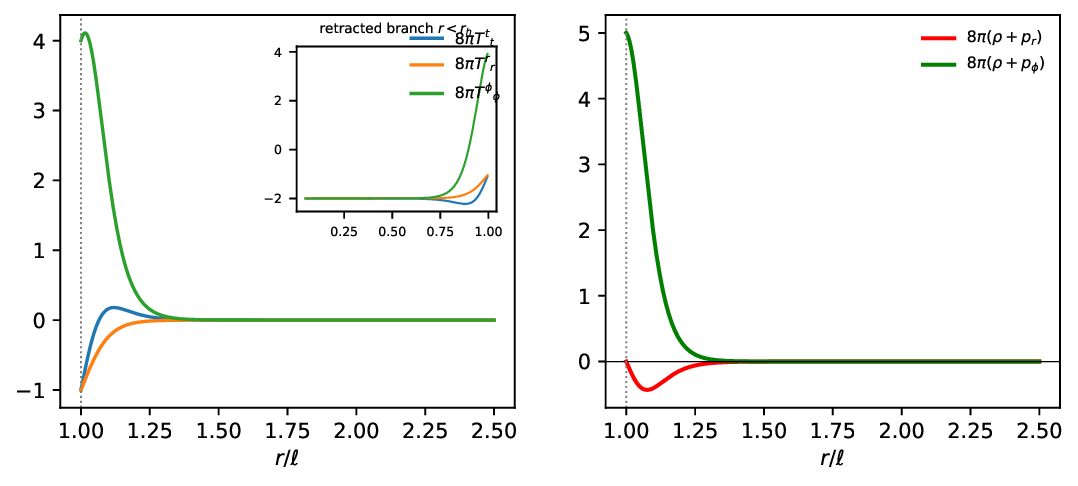}
		\caption{Left: the effective source~\eqref{eq:source} on the Lorentzian sector. Inset: the retracted branch $r<\rh$, shown for visual comparison only; it is \emph{not} part of the Lorentzian spacetime discussed in this paper. Right: the exact NEC combinations~\eqref{eq:nec}; the radial condition is violated with exponentially decaying tails, the angular one is satisfied everywhere.}
		\label{fig:source}
	\end{figure*}
	
	\section{Geodesics in the extension}\label{sec:geo}
	
	In the chart~\eqref{eq:metric} the radial equation is $\dot r^{2}=\Sd(E^{2}-F(\kappa+L^{2}/r^{2}))$, with $\kappa=1$ for timelike and $0$ for null geodesics; it has a simple zero at $\rh$. Taken at face value in that chart this suggests a turning point, but the chart is singular there and the inference is not legitimate; the question must be settled in~\eqref{eq:extension}.
	
	In the regular chart the first-order system is, for arbitrary conserved $E>0$ and $L$ (Appendix~\ref{app:A}),
	\begin{equation}
		\frac{dv}{d\tau}=\frac{E-\sqrt{E^{2}-FU}}{F},\quad
		\frac{dq}{d\tau}=-\frac{2\sqrt{E^{2}-FU}}{\beta},
		\label{eq:geo}
	\end{equation}
	with $U\equiv\kappa+L^{2}/r^{2}$. Both are finite at the throat: $dv/d\tau\to U/2E$ and $dq/d\tau\to-2E/\beta(0)$. The second limit is \emph{nonzero}, so the crossing is transversal in $q$; the apparent turning point of $r$ is an artifact of $r$ being a bad coordinate there, and the areal-radius minimum is a bounce rather than a reflection.
	
	More strongly, the Christoffel symbols of~\eqref{eq:extension} are rational in $F$, $\beta$, their first derivatives and $r$, with denominators only powers of $\beta$ and $r$ (Appendix~\ref{app:A}); since $\beta(0)=4\sqrt{\delta}\neq0$, $r(0)=\rh>0$ and $F,\beta$ are analytic, \emph{all Christoffel symbols are analytic in a neighbourhood of $q=0$}. By the Cauchy--Kovalevskaya theorem for analytic ordinary differential equations, the geodesic system therefore possesses a unique analytic solution through every point of $\Sigma$ for arbitrary initial direction, hence for every $E$ and $L$. Regularity of the crossing is thus a property of the whole family of geodesics, established analytically, not an inference from sampled trajectories. Figure~\ref{fig:extension} shows one representative case, integrated as a full system in $(v,q)$, with the norm conserved to six digits.
	
	We stress what this does \emph{not} establish. Local existence and uniqueness through $\Sigma$ is not geodesic completeness of the maximal extension. A completeness theorem requires constructing the maximal extension and its conformal diagram, treating all $E$ and $L$ including the horizon generators, and controlling the global chain of sheets---each traversal reaching the next sheet's horizon at $v\to\infty$, where a further chart is needed. That construction is not attempted here and remains deferred (Sec.~\ref{sec:scope}).
	
	\section{The throat: near-horizon geometry, entropy function, and thermodynamic status}\label{sec:throat}
	
	Introducing the proper radial distance $z$ on the exterior side, the near-throat metric takes the form $ds^{2}\simeq-Ae^{-\lambda z}dt^{2}+dz^{2}+\rh^{2}d\phi^{2}$ with $\lambda=\sqrt{F'(\rh)/\delta}$. The $(t,z)$ sector has constant curvature $-\lambda^{2}/2$: the near-throat geometry is
	\begin{equation}
		\mathrm{AdS}_{2}\times S^{1},\qquad \ell_{\mathrm{AdS}_{2}}=\sqrt{2\delta/F'(\rh)},
	\end{equation}
	the standard near-horizon geometry of a degenerate horizon. Together with $\kappa=0$ in~\eqref{eq:kappa} and the infinite proper distance, this makes the extremal character of the throat unambiguous.
	
	\subsection{Entropy from three independent routes}
	
	The throat is not a bifurcate Killing horizon, and with $\kappa=0$ there is no temperature to conjugate an entropy to. Nevertheless three independent constructions assign it the same value, and we record them because their agreement is a nontrivial consistency check on the geometry.
	
	\emph{(i) Minimal surface.} The circle $q=0$ has both null expansions vanishing: the two null directions orthogonal to it satisfy $dv=0$ and $dq/dv=F/\beta$ respectively, and along each $dr/d\lambda\propto q\to0$ at the throat, while $d^{2}r/dq^{2}=2>0$ (Appendix~\ref{app:D}). It is therefore an extremal---indeed minimal---surface, and since the areal radius attains its global minimum there, it is the minimal surface separating the two asymptotic regions of the bounce. This is precisely the covariant characterisation used in the holographic entanglement-entropy prescription~\cite{RyuTakayanagi,HRT}, and in the two-sided setting the associated quantity is the entanglement entropy between the two boundaries~\cite{MaldacenaEternal}. Its length gives
	\begin{equation}
		S_{\min}=\frac{2\pi\rh}{4G}=\frac{\pi\rh}{2G} .
		\label{eq:Smin}
	\end{equation}
	
	\emph{(ii) Wald--Noether charge.} For the Einstein--Hilbert Lagrangian the Noether-charge formula~\cite{Wald1993,IyerWald} is insensitive to the effective source, since the latter contributes no Riemann-dependent terms, and yields (Appendix~\ref{app:D})
	\begin{equation}
		S_{\mathrm{Wald}}=\frac{\mathcal{A}}{4G}=\frac{\pi\rh}{2G},
	\end{equation}
	independent of $\delta$, in agreement with~\eqref{eq:Smin}.
	
	\emph{(iii) Asymptotic symmetry and Cardy.} The exterior approaches BTZ exponentially, so the Brown--Henneaux central charge~\cite{BrownHenneaux} is the standard $c=3\ell/2G$. The Brown--York quasilocal mass~\cite{BrownYork}, computed in Appendix~\ref{app:D} directly from the metric rather than inferred, is
	\begin{equation}
		M_{\rm ql}(R)=\frac{\sqrt{F}}{4G}\Big(\frac{R}{\ell}-\sqrt{\Sd F}\Big)\Big|_{r=R}\;\xrightarrow[R\to\infty]{}\;\frac{M}{8G},
	\end{equation}
	the BTZ value, differing from it at finite $R$ only by $O(e^{-2(R-\rh)/\delta})$. Feeding $L_{0}=\bar L_{0}=M\ell/16G$ into the Cardy formula~\cite{Cardy,Strominger} gives
	\begin{equation}
		S_{\rm Cardy}=4\pi\sqrt{\frac{c\,L_{0}}{6}}=\frac{\pi\sqrt{M}\,\ell}{2G}=\frac{\pi\rh}{2G},
	\end{equation}
	again the same value. We stress that this last route is a consistency check, not a derivation: it uses only the asymptotic charges, which do not by themselves fix the dual state.
	
	\emph{What is still missing.} All three routes deliver a number; none delivers a first law. With $\kappa=0$ there is no conjugate temperature, and an extremal horizon has no bifurcation surface, which is where Wald's first-law derivation ordinarily begins. We have also not computed a Euclidean on-shell action or a Hawking state. The honest summary is therefore that the horizon-length value $\pi\rh/2G$ is robust across three independent constructions, while its thermodynamic interpretation remains open.
	
	For completeness we note why the standard variational machinery for extremal horizons is unavailable here. Sen's entropy function~\cite{Sen2005,Sen2008}, the natural tool for an AdS$_{2}$ throat, requires the near-horizon configuration to extremize an action. Ours does not: AdS$_{2}\times S^{1}$ is not an Einstein space in three dimensions for $\Lambda\neq0$ (the $\phi\phi$ component of $R_{\mu\nu}=2\Lambda g_{\mu\nu}$ would force $\Lambda v_{2}=0$), so the throat is sustained by the effective source~\eqref{eq:source}, for which no action exists; and with $J=0$ there is no Kaluza--Klein gauge field, hence no charge to Legendre transform and no nontrivial extremization. Constructing the entropy function explicitly confirms this: it is linear in the AdS$_{2}$ radius and has no critical point. The obstruction is thus the kinematic character of the model, not a property of the throat, and it would be absent for the rotating generalization, where the Kaluza--Klein vector supplies the missing structure.
	
	We record one further negative statement, because it is easy to get wrong: the $\delta\to0$ limit is \emph{discontinuous} in the surface gravity. For every $\delta>0$ one has $\kappa(\delta)=0$, whereas the pointwise limit geometry is exact BTZ with $\kappa=\rh/\ell^{2}\neq0$; thus $\lim_{\delta\to0}\kappa(\delta)\neq\kappa(\lim_{\delta\to0}g)$. The smooth family is not a deformation of BTZ that can be undone at the end of a calculation.
	
	\section{Scalar probe: mode stability and the Aretakis instability}\label{sec:scalar}
	
	\subsection{Positivity of the potential for all modes}
	
\begin{figure}[t]
		\includegraphics[width=\columnwidth]{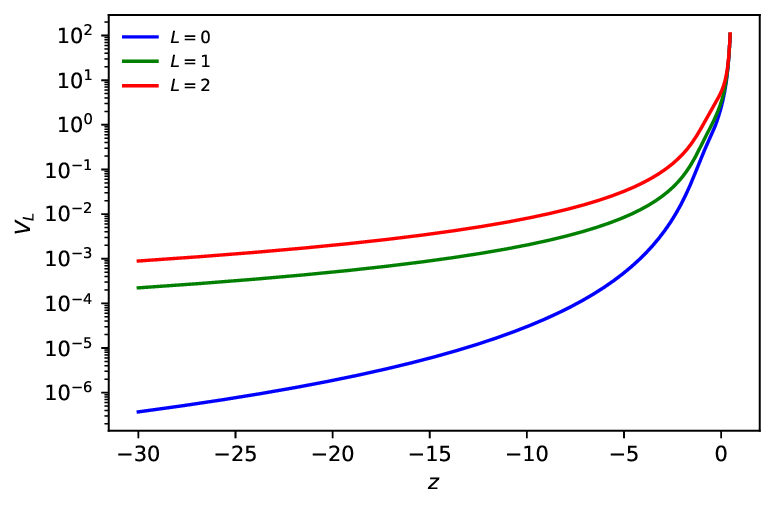}
		\caption{The effective potential~\eqref{eq:V} in the tortoise coordinate, for $L=0,1,2$: strictly positive, vanishing toward the throat and rising at the AdS boundary. Positivity is proved analytically for all $L$ in Appendix~\ref{app:C}, not merely sampled.}
		\label{fig:potential}
	\end{figure}

	For a massless scalar $\psi=R(r)e^{-i\omega t+iL\phi}$ on~\eqref{eq:metric}, with tortoise coordinate $dz=dr/(\sqrt{\Sd}F)$ and $u=\sqrt{r}R$, the radial problem takes the Schr\"odinger form $u''+(\omega^{2}-V_{L})u=0$ with (Appendix~\ref{app:C})
	\begin{equation}
		V_{L}=\frac{FL^{2}}{r^{2}}+\frac{F\big[\,rFS'+S\big(3r^{2}/\ell^{2}+M\big)\big]}{4r^{2}} .
		\label{eq:V}
	\end{equation}
	Since $F>0$, $S>0$ and $S'\geq0$ for $r>\rh$, every term in~\eqref{eq:V} is non-negative: $V_{L}>0$ strictly, for all angular modes $L$, all parameter values $M,\ell,\delta$, and any monotone transition function (Fig.~\ref{fig:potential}). This replaces the two-mode numerical check of earlier versions of this work with a proof; a numerical scan over $L\le10$ and $\delta\in[0.03,0.3]$ agrees.
	
	The consequences must be stated precisely. At the throat end $z\to-\infty$ the potential vanishes and the endpoint is in the limit-point case; at the AdS end the problem is limit-circle, so the radial operator is \emph{not} essentially self-adjoint and a self-adjoint extension must be chosen~\cite{IshibashiWald}. Choosing the Dirichlet (Friedrichs) extension, standard for normalizable AdS boundary conditions, positivity of $V_{L}$ gives a non-negative spectrum and hence the absence of exponentially growing modes. That is a statement about linear, exterior, test-scalar modes under one boundary condition; it is not a decay statement, and we do not establish local energy decay, absence of threshold resonances, or nonlinear stability.
	
	\subsection{Aretakis-type instability at the throat}
	
	Mode stability in the bulk does not preclude the instability characteristic of degenerate horizons~\cite{Aretakis1,Aretakis2,LuciettiReall}, which concerns transverse derivatives evaluated \emph{on} the horizon. In the regular chart~\eqref{eq:extension} the massless wave equation reduces to the exact form (Appendix~\ref{app:E})
	\begin{equation}
		2r\,\partial_{v}\partial_{q}\psi+r'\,\partial_{v}\psi+\partial_{q}\big(W\partial_{q}\psi\big)-\frac{\beta L^{2}}{2r}\psi=0,
		\label{eq:master}
	\end{equation}
	with $W\equiv2rF/\beta$ and $r'=dr/dq$. At $q=0$ one has $r'=0$ and, because $F$ has a double zero there, $W(0)=W'(0)=0$. Evaluating~\eqref{eq:master} on the throat for the axisymmetric mode $L=0$ therefore leaves
	\begin{equation}
		2\rh\,\partial_{v}\big(\partial_{q}\psi\big)\big|_{q=0}=0,
	\end{equation}
	so that
	\begin{equation}
		H_{0}\equiv\partial_{q}\psi\big|_{q=0}
	\end{equation}
	is \emph{exactly conserved along the throat}: the Aretakis constant of this geometry. Differentiating~\eqref{eq:master} once in $q$ and evaluating at $q=0$ gives, with $W''(0)=4\rh F'(\rh)/\beta(0)$,
	\begin{equation}
		\partial_{v}\big(\partial_{q}^{2}\psi\big)\big|_{q=0}
		=-\frac{1}{\rh}\Big[\partial_{v}\psi\big|_{q=0}+\frac{2\rh F'(\rh)}{\beta(0)}H_{0}\Big].
		\label{eq:aretakis}
	\end{equation}
	If $\psi$ settles to a constant on the throat at late advanced time, $\partial_{v}\psi|_{0}\to0$, and~\eqref{eq:aretakis} gives linear growth,
	\begin{equation}
		\partial_{q}^{2}\psi\big|_{q=0}\;\simeq\;-\frac{\rh}{\ell^{2}\sqrt{\delta}}\,H_{0}\,v ,
		\label{eq:growth}
	\end{equation}
	using $\beta(0)=4\sqrt{\delta}$ and $F'(\rh)=2\rh/\ell^{2}$. The throat therefore carries an Aretakis-type instability, with a rate that diverges as $\delta\to0$.
	
\begin{figure*}[t]
		\includegraphics[width=\textwidth]{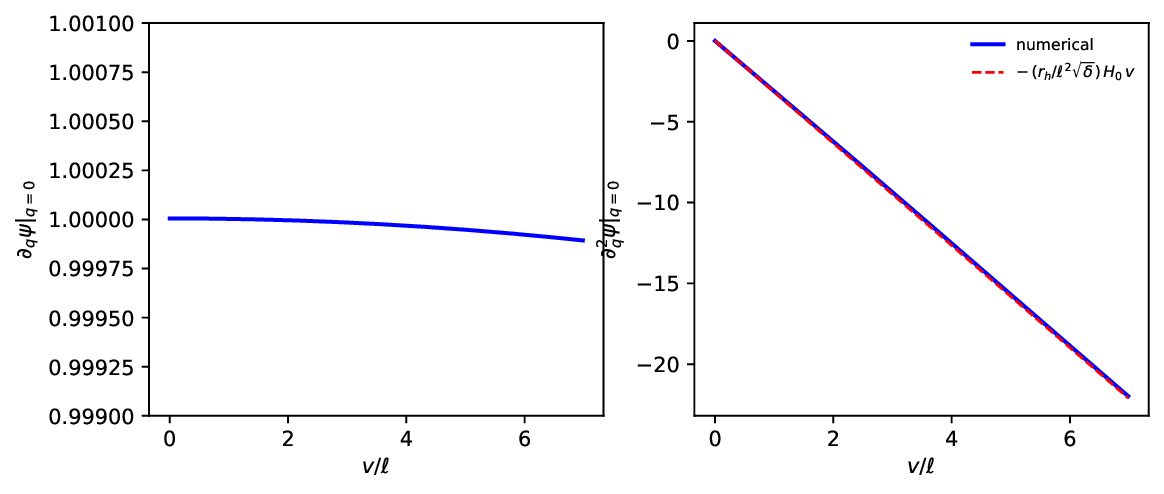}
		\caption{Numerical evolution of~\eqref{eq:master} for $L=0$ at $M=\ell=1$, $\delta=0.1$, with initial data giving $H_{0}=1$. Left: the Aretakis constant $\partial_{q}\psi|_{q=0}$, conserved to a relative drift of $8\times10^{-5}$. Right: $\partial_{q}^{2}\psi|_{q=0}$ growing linearly, against the analytic prediction~\eqref{eq:growth} (dashed).}
		\label{fig:aretakis}
	\end{figure*}

	We have confirmed both statements by direct numerical evolution of~\eqref{eq:master}, integrated with an explicit finite-difference scheme in advanced time on a uniform $q$-grid (Fig.~\ref{fig:aretakis}). With initial data $\psi(0,q)=q\,e^{-q^{2}/2(0.3)^{2}}$, so that $H_{0}=1$, the numerically measured $H_{0}$ is conserved to a relative drift of $8\times10^{-5}$ over $v=7\ell$, while the late-time slope of $\partial_{q}^{2}\psi|_{q=0}$ is $-3.170$ against the prediction $-\rh/(\ell^{2}\sqrt{\delta})=-3.162$, an agreement of $0.3\%$.
	
	Two remarks. First, the instability is of the standard degenerate-horizon type: the field itself and its first transverse derivative remain bounded on the throat, while the second and higher transverse derivatives grow polynomially, so this is a statement about the horizon, not about bulk mode stability---the two results of this section are complementary, not contradictory. Second, its presence here is a further confirmation, independent of the geometric arguments of Sec.~\ref{sec:extension}, that $\Sigma$ is a genuine degenerate Killing horizon rather than a signature-change surface: the Aretakis mechanism requires precisely a double zero of $g_{vv}$ at a regular null hypersurface. The nonlinear fate of this instability, and its relation to the trapping instabilities of horizonless ultracompact objects~\cite{Keir,CardosoPani}, are beyond our scope.
	
	\section{Scope and limitations}\label{sec:scope}
	
	We state plainly what this paper does not do.
	
	\emph{Not a signature-changing spacetime.} The construction was motivated by signature change and does not achieve it. The Lorentzian sector extends analytically to a second Lorentzian sheet; the Riemannian branch is a separate geometry, singular at its centre for the profile used here. Whether any smooth metric can implement a Lorentzian--Riemannian transition at a BTZ horizon remains open; we have shown only that two natural ansatz families do not, one because it is singular~\eqref{eq:negative} and one because its analytic extension is Lorentzian. We claim no classification theorem.
	
	\emph{Kinematic, not dynamical.} Equation~\eqref{eq:source} is the Einstein tensor of a metric written down by hand. No action, matter model, equation of state, or formation scenario is offered, and no argument is given that such a source can arise from a stable classical or quantum system.
	
	\emph{The scale $\delta$ is unexplained.} All regularity is bought with it: the transition curvature~\eqref{eq:inv} scales as $1/\delta$ and $1/\delta^{2}$, and the Aretakis growth rate~\eqref{eq:growth} as $1/\sqrt{\delta}$. Nothing here fixes $\delta$, and the model inherits the standard objection to regular-black-hole ans\"atze: if $\delta$ is of quantum-gravitational size, the validity of a purely classical treatment at the shell may be challenged, while a macroscopic $\delta$ predicts an unexplained new scale with observational consequences (absence of Hawking radiation, throat echoes, shell energy-condition violation) that we have not analysed.
	
	\emph{Deferred.} Geometrically, the maximal extension and conformal diagram for all $E,L$ remain to be constructed, geodesic completeness proved as a theorem including horizon generators and the global chain of sheets, and the nonlinear development of the Aretakis and trapping instabilities studied. Thermodynamically, the Cardy agreement of Sec.~\ref{sec:throat} is a consistency check rather than a microscopic derivation, since the dual state has not been identified; a first law at $\kappa=0$, a Euclidean on-shell action, and the resulting CFT interpretation of removing the horizon and inserting exotic matter near it are all open. More broadly, rotating and charged generalizations -- where a Kaluza--Klein gauge field would make the entropy function nondegenerate -- a rigorous distributional treatment of the $\delta\to0$ junction, for which ordinary Israel theory is inapplicable because the normal degenerates, and the four-dimensional analogue, where the same substitution leaves a curvature singularity at the centre of the Riemannian branch, are left for future work.
	
	\section{Conclusions}\label{sec:concl}
	
	Inserting a smooth transition function into $g^{rr}$ of the BTZ metric does not produce a signature-changing spacetime. It produces a black bounce whose throat sits exactly at the would-be horizon and is therefore degenerate: a regular null hypersurface with vanishing surface gravity, AdS$_{2}\times S^{1}$ near-throat geometry, and a second isometric exterior beyond it. Curvature invariants are finite there and given in closed form; the effective source is explicit; the radial null energy condition is violated in the transition shell, with tails, by an amount we quantify invariantly; the geodesic system is analytic at the throat for the entire family of conserved charges; and the scalar effective potential is strictly positive for every mode, which we prove rather than sample.
	
	Two results sharpen the extremal interpretation. The throat circle is the minimal surface separating the two exteriors, with vanishing null expansions, and the entropy $\pi\rh/2G$ that its length assigns is reproduced independently by the Wald--Noether charge and by a Cardy estimate built on the Brown--Henneaux central charge and the computed Brown--York mass---three concordant determinations of a quantity for which, $\kappa$ being zero, no first law is yet available. And the throat carries an Aretakis-type instability, with an exactly conserved $\partial_{q}\psi|_{q=0}$ and linearly growing $\partial_{q}^{2}\psi|_{q=0}$ at a rate we derive analytically and confirm numerically. Alongside these we record the negative result that the alternative ansatz, smoothing $g_{tt}$ as in the Lorentzian--Euclidean Schwarzschild proposal, is singular at the horizon for every finite smoothing width.

	\appendix
	
	\section{The regular chart and the extension}\label{app:A}
	
	\subsection{Analyticity of $\sigma(q)$}
	
	Write $\tanh y=y\,g(y)$, where
	\begin{equation}
		g(y)=\frac{\tanh y}{y}=1-\frac{y^{2}}{3}+\frac{2y^{4}}{15}-\cdots
	\end{equation}
	is even, analytic on $|y|<\pi/2$ and $g(0)=1$. With $r-\rh=q^{2}$ and $y=q^{2}/\delta$,
	\begin{equation}
		\Sd=\tanh\frac{q^{2}}{\delta}=\frac{q^{2}}{\delta}\,g\!\Big(\frac{q^{2}}{\delta}\Big),
	\end{equation}
	so the signed square root is
	\begin{equation}
		\sigma(q)=q\,\sqrt{\frac{g(q^{2}/\delta)}{\delta}},
		\label{eq:sigma}
	\end{equation}
	which is manifestly odd, and analytic near $q=0$ because $g$ is analytic and positive there, so its square root is analytic. Expanding,
	\begin{equation}
		\sigma(q)=\frac{q}{\sqrt{\delta}}\Big(1-\frac{q^{4}}{6\delta^{2}}+O(q^{8})\Big),
	\end{equation}
	which at $\delta=1/10$ gives $\sigma=\sqrt{10}\,q-\tfrac{50\sqrt{10}}{3}q^{5}+O(q^{9})$, the series quoted in Sec.~\ref{sec:extension}. Note $\sigma^{2}=\Sd$ exactly, and $\sigma<0$ for $q<0$: this sign choice is what distinguishes the extension from a mere double cover.
	
	\subsection{The extension metric}
	
	Under~\eqref{eq:coordchange}, $dt=dv-dr/(F\sigma)$ and $dr=2q\,dq$, so
	\begin{align}
		-F\,dt^{2}&=-F\,dv^{2}+\frac{2}{\sigma}\,dv\,dr-\frac{dr^{2}}{F\sigma^{2}},\nonumber\\
		\frac{dr^{2}}{\Sd F}&=\frac{dr^{2}}{\sigma^{2}F}.
	\end{align}
	The $dr^{2}$ terms cancel identically---this is the point of the construction---leaving
	\begin{equation}
		ds^{2}=-F\,dv^{2}+\frac{2}{\sigma}\,dv\,dr+r^{2}d\phi^{2}
		=-F\,dv^{2}+\beta\,dv\,dq+r^{2}d\phi^{2},
	\end{equation}
	with
	\begin{equation}
		\beta=\frac{2}{\sigma}\frac{dr}{dq}=\frac{4q}{\sigma}=4\sqrt{\frac{\delta}{g(q^{2}/\delta)}},
	\end{equation}
	where the last equality uses~\eqref{eq:sigma}. Since $g$ is even in $q$, $\beta$ is an even, analytic, strictly positive function with $\beta(0)=4\sqrt{\delta}$.
	
	\subsection{Determinant and the null character of $\Sigma$}
	
	In the ordering $(v,q,\phi)$,
	\begin{equation}
		g_{\mu\nu}=\begin{pmatrix}-F&\beta/2&0\\ \beta/2&0&0\\ 0&0&r^{2}\end{pmatrix},
	\end{equation}
	whose $(v,q)$ block has determinant
	\begin{equation}
		(-F)(0)-\Big(\frac{\beta}{2}\Big)^{2}=-\frac{\beta^{2}}{4},
	\end{equation}
	independent of $F$ and equal to $-4\delta<0$ at $q=0$. Hence $\det g=-\beta^{2}r^{2}/4$ is finite and negative there: the metric is Lorentzian and non-degenerate. Inverting,
	\begin{equation}
		g^{vv}=0,\qquad g^{vq}=\frac{2}{\beta},\qquad g^{qq}=\frac{4F}{\beta^{2}},\qquad g^{\phi\phi}=\frac{1}{r^{2}} .
		\label{eq:inverse}
	\end{equation}
	Since $F(\rh)=0$, $g^{qq}$ vanishes at $q=0$, i.e.\ $\nabla_{\mu}q$ is null there: $\Sigma$ is a null hypersurface.
	
	\subsection{Christoffel symbols and geodesic regularity}
	
	With $F=F(q)$, $\beta=\beta(q)$ and $r=\rh+q^{2}$, the nonvanishing symbols are
	\begin{align}
		\Gamma^{v}{}_{vv}&=\frac{F'}{\beta}, &
		\Gamma^{v}{}_{\phi\phi}&=-\frac{4qr}{\beta},\nonumber\\
		\Gamma^{q}{}_{vv}&=\frac{2FF'}{\beta^{2}}, &
		\Gamma^{q}{}_{vq}&=-\frac{F'}{\beta},\nonumber\\
		\Gamma^{q}{}_{qq}&=\frac{\beta'}{\beta}, &
		\Gamma^{q}{}_{\phi\phi}&=-\frac{8qrF}{\beta^{2}},\nonumber\\
		\Gamma^{\phi}{}_{q\phi}&=\frac{2q}{r}. &&
	\end{align}
	Every symbol is a rational expression in $F,\beta,F',\beta',q,r$ whose only denominators are powers of $\beta$ and $r$. Since $\beta(0)=4\sqrt{\delta}\neq0$ and $r(0)=\rh>0$, and $F,\beta$ are analytic in $q$ near $0$, all Christoffel symbols are analytic in a neighbourhood of $q=0$. The geodesic equations $\ddot x^{\mu}+\Gamma^{\mu}{}_{\alpha\beta}\dot x^{\alpha}\dot x^{\beta}=0$ therefore form an analytic ODE system, and by Cauchy--Kovalevskaya admit a unique analytic solution through any point of $\Sigma$ with any initial direction.
	
	For the first-order form, the conserved quantities are $E=F\dot v-\tfrac12\beta\dot q$ and $L=r^{2}\dot\phi$, and the normalization $g_{\mu\nu}\dot x^{\mu}\dot x^{\nu}=-\kappa$ reduces to $F\dot v^{2}-2E\dot v+U=0$ with $U=\kappa+L^{2}/r^{2}$, giving~\eqref{eq:geo}. As $F\to0$,
	\begin{equation}
		\dot v=\frac{E-\sqrt{E^{2}-FU}}{F}=\frac{U}{2E}+\frac{FU^{2}}{8E^{3}}+O(F^{2}),
	\end{equation}
	finite, and $\dot q\to-2E/\beta(0)\neq0$ for $E>0$: the crossing is transversal. (Numerically at $E=1.3$, $L=0.7$, $\kappa=1$, $f_{1}=2$: $\dot v=0.573065,\,0.573077,\,0.573077$ at $q=10^{-2},10^{-3},10^{-4}$ against $U/2E=0.573077$.)
	
	\section{Curvature and the effective source}\label{app:B}
	
	For $ds^{2}=-F\,dt^{2}+dr^{2}/(SF)+r^{2}d\phi^{2}$ with generic smooth $S(r)$, the nonvanishing Christoffel symbols are
	\begin{align}
		\Gamma^{t}{}_{tr}&=\frac{F'}{2F}, &
		\Gamma^{r}{}_{tt}&=\frac{SFF'}{2},\nonumber\\
		\Gamma^{r}{}_{rr}&=-\frac{(SF)'}{2SF}, &
		\Gamma^{r}{}_{\phi\phi}&=-rSF,\nonumber\\
		\Gamma^{\phi}{}_{r\phi}&=\frac{1}{r}. &&
	\end{align}
	Computing $R_{\mu\nu}=\partial_{\alpha}\Gamma^{\alpha}{}_{\mu\nu}-\partial_{\nu}\Gamma^{\alpha}{}_{\mu\alpha}+\Gamma^{\alpha}{}_{\alpha\lambda}\Gamma^{\lambda}{}_{\mu\nu}-\Gamma^{\alpha}{}_{\mu\lambda}\Gamma^{\lambda}{}_{\alpha\nu}$ and raising one index,
	\begin{align}
		R^{t}{}_{t}&=-SF''-\frac{S'F'}{4},\nonumber\\
		R^{r}{}_{r}&=-SF''-\frac{S'F'}{4}-\frac{FS'}{2r},\label{eq:mixedRicci}\\
		R^{\phi}{}_{\phi}&=-\frac{SF''}{2}-\frac{(SF)'}{2r},\nonumber
	\end{align}
	with $F=-M+r^{2}/\ell^{2}$, so that $F'=2r/\ell^{2}$ and $F''=2/\ell^{2}$. As a check, $S\equiv1$ gives $R^{\mu}{}_{\nu}=-(2/\ell^{2})\delta^{\mu}{}_{\nu}$, the BTZ values; and on $\Sigma$, where $S=F=0$, Eq.~\eqref{eq:mixedRicci} gives $R^{t}{}_{t}=R^{r}{}_{r}=-S'(\rh)F'(\rh)/4$ and $R^{\phi}{}_{\phi}=0$, as quoted in Sec.~\ref{sec:curvature}. Summing, the Ricci scalar is
	\begin{equation}
		R=-\frac{6S}{\ell^{2}}-\frac{FS'}{r}-\frac{rS'}{\ell^{2}} ,
		\label{eq:Rscalar}
	\end{equation}
	which reduces to $-6/\ell^{2}$ when $S\equiv1$ (BTZ) and, on $\Sigma$ where $S=0$ and $F=0$, to $-\rh S'(\rh)/\ell^{2}=-\rh/(\ell^{2}\delta)$, the value quoted in~\eqref{eq:inv}.
	
	In three dimensions the Weyl tensor vanishes identically, so the Riemann tensor is determined algebraically by the Ricci tensor,
	\begin{equation}
		R_{\mu\nu\rho\sigma}=g_{\mu\rho}R_{\nu\sigma}+g_{\nu\sigma}R_{\mu\rho}-g_{\mu\sigma}R_{\nu\rho}-g_{\nu\rho}R_{\mu\sigma}-\frac{R}{2}\big(g_{\mu\rho}g_{\nu\sigma}-g_{\mu\sigma}g_{\nu\rho}\big),
	\end{equation}
	and contracting gives the identity used in the text,
	\begin{equation}
		K\equiv R_{\mu\nu\rho\sigma}R^{\mu\nu\rho\sigma}=4R_{\mu\nu}R^{\mu\nu}-R^{2} .
	\end{equation}
	Evaluating on $\Sigma$ with $R^{t}{}_{t}=R^{r}{}_{r}=-F'(\rh)S'(\rh)/4$ and $R^{\phi}{}_{\phi}=0$ gives $K(\Sigma)=\rh^{2}/(\ell^{4}\delta^{2})$.
	
	Finally, with $G^{\mu}{}_{\nu}=R^{\mu}{}_{\nu}-\tfrac12R\,\delta^{\mu}{}_{\nu}$ and $8\pi T^{\mu}{}_{\nu}=G^{\mu}{}_{\nu}+\Lambda\delta^{\mu}{}_{\nu}$, $\Lambda=-1/\ell^{2}$, substitution of the above yields~\eqref{eq:source}. The three-dimensional trace relation $G^{\mu}{}_{\mu}=-R/2$ provides a useful check: summing~\eqref{eq:source} gives $8\pi T^{\mu}{}_{\mu}=-R/2+3\Lambda$, consistent with~\eqref{eq:Rscalar}.
	
	\section{The scalar potential and its positivity}\label{app:C}
	
	For $\Box\psi=0$ on~\eqref{eq:metric}, with $\sqrt{-g}=r/\sqrt{S}$ and $g^{rr}=SF$, $g^{tt}=-1/F$, separation $\psi=R(r)e^{-i\omega t+iL\phi}$ gives
	\begin{equation}
		\frac{\sqrt{S}}{r}\frac{d}{dr}\Big[r\sqrt{S}\,F\,R'\Big]+\Big(\frac{\omega^{2}}{F}-\frac{L^{2}}{r^{2}}\Big)R=0 .
		\label{eq:radial}
	\end{equation}
	Define $A\equiv\sqrt{S}F$ and the tortoise coordinate $dz=dr/A$, so $\partial_{z}=A\partial_{r}$. Multiplying~\eqref{eq:radial} by $F$ and using $A\,\partial_{r}=\partial_{z}$,
	\begin{equation}
		\frac{1}{r}\partial_{z}\big(r\,\partial_{z}R\big)+\Big(\omega^{2}-\frac{FL^{2}}{r^{2}}\Big)R=0 .
	\end{equation}
	The substitution $u=\sqrt{r}\,R$ removes the first-derivative term: writing $R=u/\sqrt{r}$,
	\begin{equation}
		\frac{1}{r}\partial_{z}\big(r\,\partial_{z}(r^{-1/2}u)\big)=r^{-1/2}\Big[\partial_{z}^{2}u-\frac{u}{\sqrt{r}}\partial_{z}^{2}\sqrt{r}\Big],
	\end{equation}
	so that $u''+(\omega^{2}-V_{L})u=0$ with
	\begin{equation}
		V_{L}=\frac{FL^{2}}{r^{2}}+\frac{1}{\sqrt{r}}\,\partial_{z}^{2}\sqrt{r} .
	\end{equation}
	Evaluating the second term with $\partial_{z}=A\partial_{r}$ and $A=\sqrt{S}F$,
	\begin{align}
		\partial_{z}\sqrt{r}&=\frac{A}{2\sqrt{r}},\nonumber\\
		\partial_{z}^{2}\sqrt{r}&=A\,\partial_{r}\Big(\frac{A}{2\sqrt{r}}\Big)
		=\frac{A A'}{2\sqrt{r}}-\frac{A^{2}}{4r^{3/2}},
	\end{align}
	and using $A A'=\tfrac12(A^{2})'=\tfrac12(SF^{2})'=\tfrac12\big(S'F^{2}+2SFF'\big)$,
	\begin{equation}
		\frac{1}{\sqrt{r}}\partial_{z}^{2}\sqrt{r}
		=\frac{S'F^{2}+2SFF'}{4r}-\frac{SF^{2}}{4r^{2}} .
	\end{equation}
	With $F'=2r/\ell^{2}$ and $F=-M+r^{2}/\ell^{2}$, the last two terms combine as
	\begin{equation}
		\frac{F}{4r^{2}}\Big[rFS'+S\big(4r^{2}/\ell^{2}-F\big)\Big]
		=\frac{F}{4r^{2}}\Big[rFS'+S\Big(\frac{3r^{2}}{\ell^{2}}+M\Big)\Big],
	\end{equation}
	which is~\eqref{eq:V}.
	
	\emph{Positivity.} For $r>\rh$ we have, term by term: $F=-M+r^{2}/\ell^{2}>0$ since $r>\rh=\sqrt{M}\ell$; $S=\Sd>0$ since $r>\rh$; $S'\geq0$ for any monotone non-decreasing transition function; $3r^{2}/\ell^{2}+M>0$ trivially; and $L^{2}\geq0$. Hence
	\begin{equation}
		rFS'\geq0,\qquad S\Big(\frac{3r^{2}}{\ell^{2}}+M\Big)>0,\qquad \frac{FL^{2}}{r^{2}}\geq0,
	\end{equation}
	so every contribution to $V_{L}$ in~\eqref{eq:V} is non-negative and the second is strictly positive. Therefore $V_{L}>0$ for all $r>\rh$, all $L\in\mathbb{Z}$, all $M,\ell>0$ and $\delta>0$, and any monotone $S$. No property of $\tanh$ beyond monotonicity and positivity on $r>\rh$ was used. As $r\to\rh^{+}$, $F\to0$ and $V_{L}\to0$; as $r\to\infty$, $S\to1$ and $V_{L}\simeq3r^{2}/(4\ell^{4})\to\infty$, the AdS wall.
	
	\section{Entropy and quasilocal mass}\label{app:D}
	
	\subsection{The throat is a minimal surface}
	
	Null directions orthogonal to a circle $q=\mathrm{const}$ in the chart~\eqref{eq:extension} satisfy $-F\,dv^{2}+\beta\,dv\,dq=0$, i.e.
	\begin{equation}
		dv=0 \quad\text{(ingoing)},\qquad \frac{dq}{dv}=\frac{F}{\beta}\quad\text{(outgoing)} .
	\end{equation}
	The expansion of the circle along either congruence is $\theta=r^{-1}dr/d\lambda$ with $r=\rh+q^{2}$, so $dr=2q\,dq$ and
	\begin{equation}
		\theta_{\rm in}\propto 2q,\qquad \theta_{\rm out}\propto 2q\,\frac{F}{\beta},
	\end{equation}
	both of which vanish at $q=0$. The circle $\Sigma$ therefore has vanishing null expansions---the covariant characterisation of an extremal surface~\cite{HRT}---and since
	\begin{equation}
		\frac{d^{2}r}{dq^{2}}=2>0,
	\end{equation}
	the areal radius attains a strict minimum there, so the surface is minimal rather than merely extremal. Its length is $\mathcal{A}=2\pi\rh$, giving Eq.~\eqref{eq:Smin}. Equivalently, on a static slice the circle $r=\mathrm{const}$ has geodesic curvature $k_{g}=\sqrt{\Sd F}/r$, which vanishes at $r=\rh$: the throat circle is a closed geodesic of the spatial geometry. (On such a slice it lies at infinite proper distance, the usual feature of an extremal throat; the expansion computation above is performed in the regular chart and is free of that limitation.)
	
	\subsection{Wald--Noether charge}
	
	The Wald entropy~\cite{Wald1993,IyerWald} is
	\begin{equation}
		S_{\mathrm{Wald}}=-2\pi\oint\frac{\partial\mathcal{L}}{\partial R_{\mu\nu\rho\sigma}}\,\epsilon_{\mu\nu}\epsilon_{\rho\sigma}\,d\mathcal{A},
	\end{equation}
	with $\epsilon_{\mu\nu}$ the binormal normalized by $\epsilon_{\mu\nu}\epsilon^{\mu\nu}=-2$. For $\mathcal{L}=(R-2\Lambda)/16\pi G$,
	\begin{equation}
		\frac{\partial\mathcal{L}}{\partial R_{\mu\nu\rho\sigma}}=\frac{1}{32\pi G}\big(g^{\mu\rho}g^{\nu\sigma}-g^{\mu\sigma}g^{\nu\rho}\big),
	\end{equation}
	whose contraction with $\epsilon_{\mu\nu}\epsilon_{\rho\sigma}$ gives $-2/32\pi G$, so the integrand is $1/4G$ per unit length and
	\begin{equation}
		S_{\mathrm{Wald}}=\frac{\mathcal{A}}{4G}=\frac{2\pi\rh}{4G}=\frac{\pi\rh}{2G} .
	\end{equation}
	The cosmological and effective-source terms contribute nothing, having no Riemann dependence; the result is therefore independent of $\delta$ and of the details of the transition function. Note that the derivation of the first law from this charge assumes a bifurcation surface, which a degenerate horizon does not possess; we use the formula here only as a definition of the Noether charge on the throat cross-section.
	
	\subsection{Brown--York quasilocal mass}
	
	For $ds^{2}=-N^{2}dt^{2}+dr^{2}/f+r^{2}d\phi^{2}$ with $N^{2}=F$ and $f=\Sd F$, the boundary circle at $r=R$ on a static slice has outward unit normal $n=\sqrt{f}\,\partial_{r}$ and extrinsic curvature $k=\sqrt{f}/R$. The Brown--York energy~\cite{BrownYork} relative to a background with $k_{0}=1/\ell$ (pure AdS$_{3}$) is $E=(R/4G)(k_{0}-k)$, and the conserved mass is obtained by multiplying by the lapse,
	\begin{equation}
		M_{\rm ql}(R)=N(R)\,\frac{R}{4G}\Big(\frac{1}{\ell}-\frac{\sqrt{\Sd F}}{R}\Big)
		=\frac{\sqrt{F}}{4G}\Big(\frac{R}{\ell}-\sqrt{\Sd F}\Big),
	\end{equation}
	all quantities evaluated at $r=R$. Since $\Sd\to1$ exponentially,
	\begin{equation}
		M_{\rm ql}(R)=\frac{M}{8G}+O\!\big(R^{-2}\big)+O\!\big(e^{-2(R-\rh)/\delta}\big)\;\xrightarrow[R\to\infty]{}\;\frac{M}{8G},
	\end{equation}
	the $R^{-2}$ terms being the usual background-subtraction corrections present already for BTZ. As a check, $M_{\rm ql}(R)$ was evaluated by direct substitution into the closed form above at a sequence of radii; numerically at $M=\ell=G=1$, $\delta=0.1$: $M_{\rm ql}=0.11603,\,0.12372,\,0.124686,\,0.1249875$ at $R/\ell=2,5,10,50$, against $M/8G=0.125$; at $R=2\ell$ the value agrees with that of exact BTZ to nine digits, the difference being the exponentially small $\Sd-1$. The asymptotic charges are thus those of BTZ with mass parameter $M$, computed rather than assumed.
	
	\subsection{Cardy consistency check}
	
	With Brown--Henneaux~\cite{BrownHenneaux} $c=3\ell/2G$ and, for a non-rotating configuration of mass $M_{\rm ADM}=M/8G$, $L_{0}=\bar L_{0}=M_{\rm ADM}\ell/2=M\ell/16G$, the Cardy formula~\cite{Cardy,Strominger} gives
	\begin{equation}
		S_{\rm Cardy}=4\pi\sqrt{\frac{c\,L_{0}}{6}}
		=4\pi\sqrt{\frac{1}{6}\cdot\frac{3\ell}{2G}\cdot\frac{M\ell}{16G}}
		=\frac{\pi\sqrt{M}\,\ell}{2G}=\frac{\pi\rh}{2G},
	\end{equation}
	coinciding with $S_{\min}$ and $S_{\mathrm{Wald}}$. As noted in the main text this is a consistency check rather than a derivation: it uses only the asymptotic charges, and the dual state is not determined here.
	
	\section{Aretakis-type analysis at the throat}\label{app:E}
	
	\subsection{The master equation}
	
	Using the inverse metric~\eqref{eq:inverse} and $\sqrt{-g}=\beta r/2$, the massless wave operator in the chart~\eqref{eq:extension} is
	\begin{equation}
		\Box\psi=\frac{2}{\beta r}\Big[\partial_{v}\big(\tfrac{\beta r}{2}g^{vq}\partial_{q}\psi\big)+\partial_{q}\big(\tfrac{\beta r}{2}(g^{qv}\partial_{v}\psi+g^{qq}\partial_{q}\psi)\big)\Big]+g^{\phi\phi}\partial_{\phi}^{2}\psi .
	\end{equation}
	With $g^{vq}=2/\beta$ and $g^{qq}=4F/\beta^{2}$ the bracket becomes
	\begin{equation}
		r\,\partial_{v}\partial_{q}\psi+\partial_{q}\big(r\partial_{v}\psi\big)+\partial_{q}\Big(\frac{2rF}{\beta}\partial_{q}\psi\Big),
	\end{equation}
	so that, for $\psi\propto e^{iL\phi}$ and multiplying by $\beta r/2$,
	\begin{equation}
		2r\,\partial_{v}\partial_{q}\psi+r'\,\partial_{v}\psi+\partial_{q}\big(W\partial_{q}\psi\big)-\frac{\beta L^{2}}{2r}\psi=0,
		\qquad W\equiv\frac{2rF}{\beta},
	\end{equation}
	which is~\eqref{eq:master}. (We verified this reduction symbolically against a direct computation of $\sqrt{-g}\,\Box\psi$.)
	
	\subsection{The conserved transverse derivative}
	
	Near $q=0$ write $F=f_{1}q^{2}+f_{2}q^{4}+O(q^{6})$ with $f_{1}=F'(\rh)>0$ --- the double zero being exactly the statement that the horizon is degenerate --- and $\beta=\beta_{0}+\beta_{2}q^{2}+O(q^{4})$ with $\beta_{0}=4\sqrt{\delta}$. Then
	\begin{equation}
		W=\frac{2rF}{\beta}=\frac{2\rh f_{1}}{\beta_{0}}q^{2}+O(q^{4}),
	\end{equation}
	so
	\begin{equation}
		W(0)=0,\qquad W'(0)=0,\qquad W''(0)=\frac{4\rh f_{1}}{\beta_{0}} .
		\label{eq:Wdata}
	\end{equation}
	Also $r'=2q$ vanishes at $q=0$. Evaluating the master equation~\eqref{eq:master} at $q=0$ for $L=0$, the terms $r'\partial_{v}\psi$, $W'\partial_{q}\psi$ and $W\partial_{q}^{2}\psi$ all vanish, leaving
	\begin{equation}
		2\rh\,\partial_{v}\big(\partial_{q}\psi\big)\big|_{q=0}=0 .
	\end{equation}
	Hence $H_{0}=\partial_{q}\psi|_{q=0}$ is conserved along the throat, for arbitrary (smooth, $L=0$) initial data. For $L\neq0$ the same evaluation gives $\partial_{v}(\partial_{q}\psi)|_{0}=\beta_{0}L^{2}\psi|_{0}/(4\rh^{2})$, so $\partial_{q}\psi|_{0}$ still tends to a constant whenever $\psi|_{0}$ decays.
	
	\subsection{Linear growth of the second transverse derivative}
	
	Differentiating~\eqref{eq:master} (with $L=0$) once with respect to $q$,
	\begin{align}
		2r'\partial_{v}\partial_{q}\psi&+2r\,\partial_{v}\partial_{q}^{2}\psi+r''\partial_{v}\psi+r'\partial_{v}\partial_{q}\psi\nonumber\\
		&+W''\partial_{q}\psi+2W'\partial_{q}^{2}\psi+W\partial_{q}^{3}\psi=0 .
	\end{align}
	At $q=0$, using $r'=0$, $r''=2$ and~\eqref{eq:Wdata}, this reduces to
	\begin{equation}
		2\rh\,\partial_{v}\big(\partial_{q}^{2}\psi\big)\big|_{0}+2\,\partial_{v}\psi\big|_{0}+\frac{4\rh f_{1}}{\beta_{0}}H_{0}=0,
	\end{equation}
	i.e.\ Eq.~\eqref{eq:aretakis}. When $\psi|_{0}$ settles to a constant, $\partial_{v}\psi|_{0}\to0$ and
	\begin{equation}
		\partial_{v}\big(\partial_{q}^{2}\psi\big)\big|_{0}\;\longrightarrow\;-\frac{2f_{1}}{\beta_{0}}H_{0}
		=-\frac{\rh}{\ell^{2}\sqrt{\delta}}H_{0},
	\end{equation}
	using $f_{1}=2\rh/\ell^{2}$ and $\beta_{0}=4\sqrt{\delta}$, which integrates to the linear growth~\eqref{eq:growth}. Higher transverse derivatives may be treated by the same recursion, differentiating $n$ times and using $W^{(k)}(0)=0$ for $k<2$; as for extremal Reissner--Nordstr\"om~\cite{Aretakis1,Aretakis2}, one expects $\partial_{q}^{n}\psi|_{0}\sim v^{\,n-1}$.
	
	\subsection{Numerical confirmation}
	
	Equation~\eqref{eq:master} with $L=0$ was integrated on $q\in[0,1.1]$ with $1600$ points and $dv=1.2\times10^{-4}$ to $v=7\ell$, at $M=\ell=\rh=1$, $\delta=0.1$. Writing the equation as $\partial_{q}(2\sqrt{r}\,u)=-r^{-1/2}\partial_{q}(W\partial_{q}\psi)$ for $u=\partial_{v}\psi$ and integrating inward from the outer boundary avoids differentiating the flux twice and is numerically stable. With initial data $\psi(0,q)=q\exp[-q^{2}/2(0.3)^{2}]$, giving $H_{0}=1$:
	\begin{itemize}
		\item $\partial_{q}\psi|_{q=0}$ remains $1$ to a relative drift of $8\times10^{-5}$ over the full run;
		\item the late-time slope of $\partial_{q}^{2}\psi|_{q=0}$ is $-3.170$, against the analytic prediction $-\rh/(\ell^{2}\sqrt{\delta})=-3.162$, an agreement of $0.3\%$.
	\end{itemize}
	Both are shown in Fig.~\ref{fig:aretakis}.
	
	\bibliographystyle{apsrev4-2}
	\bibliography{references}
	
\end{document}